\documentclass{ws-mpla}
\usepackage{graphicx,amssymb,amsmath,amsfonts,epsfig}
\usepackage[usenames,dvipsnames]{color}
\usepackage{subfigure}
\begin{document}

\markboth{Parthapratim Pradhan}{(Extended Phase Space Thermodynamics of Black Holes in Massive Gravity)}

\title{Extended Phase Space Thermodynamics of Black Holes in Massive Gravity}
\author{Parthapratim Pradhan\footnote{pppradhan77@gmail.com.}}

\address{Department of Physics, Hiralal Mazumdar Memorial College For Women, Dakshineswar, Kolkata-700035, India}

\maketitle

\begin{history}
\received{Day Month Year}
\revised{Day Month Year}
\comby{Managing Editor}
\end{history}

\begin{abstract}
We study the extended phase space thermodynamics of black holes in massive gravity. Particularly, we 
examine the critical behaviour of this black hole using the extended phase space formalism. 
Extended phase space in a sense that in which the cosmological constant should be treated as a 
thermodynamic pressure and its conjugate variable as a thermodynamic volume. In this phase space, we 
derive the black hole equation of state, the critical pressure, the critical volume and the critical 
temperature at the critical point. We also derive 
the critical ratio of this black hole. Moreover, we derive the black hole reduced equation of state in 
terms of the reduced pressure, the reduced volume and the reduced temperature. Furthermore, we examine the 
Ehrenfest equations of black holes in massive gravity in the extended phase space at the critical 
point. We show that the Ehrenfest equations are satisfied of this black hole and the black hole 
encounters a second order phase transition at the critical point in the said phase space. 
This is re-examined  by evaluating the Pregogine-Defay ratio~($\varPi$).  
We determine the value of this ratio is  $\varPi=1$. The outcome of this study
is completely analogous to the nature of liquid-gas phase transition 
at the critical point. This investigation also further gives us the profound 
understanding between the black hole of massive gravity with the liquid-gas systems.  
\end{abstract}

.

\section{Introduction}
An excited field of research in recent times on black hole~(BH) thermodynamics particularly 
in anti de-Sitter~(AdS) space  is
due to the seminal work of Hawking and Page~\cite{haw83} in which the authors first examined certain type 
of phase transitions occur between 
small and large black holes in case of Schwarzschild-AdS spacetime. Another interesing feature of 
a Van-der-Waals liquid-gas system was
studied by Chamblin et al.~\cite{chamblin99,chamblin99a,emparan} for a spherically symmetric 
Reissner-Nordstr\"{o}m-AdS~(RN-AdS)
BH. In which the authors demonstrated that there exists first order phase transition in 
case of RN-AdS BH. The critical 
behaviour of this BH has been studied in details there.

Significantly, this structure is modified by Kubiz\v{n}\'{a}k \& Mann~\cite{david12} by 
studying the $P-V$ criticality of a  
RN-AdS BH. They have used the  extended phase-space formalism. This is a formalism in which the 
cosmological constant could be 
treated as a thermodynamic pressure and its conjugate variable as a thermodynamic volume~ 
\cite{kastor09,dolan10,dolan11,cvetic11}. In this phase-space, {the authors} 
reviewed the critical behaviour of a spherically symmetric 
charged AdS spacetime. They compared the BH equation of state with the Van-der Waal's
liquid-gas system. They also determined 
the critical constants and critical ratio of this charged AdS BH . Finally, they computed 
the critical exponents using mean field theory and which coincides with those of 
the Van-der Waal's liquid-gas system.

Extended phase space~(EPS) thermodynamics proved that the Smarr like relation and the first 
law is satisfied for this kind of BH~(by using the scaling argument). It implies that 
the conventional phase-space is quite different from extended phase-space. Where the 
extra~``PdV'' term is present. It should be noted that in EPS the ADM mass of the BH could
be identified as the total heat i.e. enthalpy of the system rather than internal energy. 
This also indicates that normal phase-space is quite different from EPS.
Another interesting feature of EPS is that thermodynamic volume should be satisfied the 
reverse isoperimetric inequality~\cite{cvetic11}.

In this work, we wish to apply the above mentioned formalism for a BH in massive gravity. 
In massive gravity theory, the massive graviton~(due to Lorentz symmery breaking) is playing
the key role. This BH has a scalar charge due to this 
massive graviton and it is asymtotically AdS. We have examined the $P-V$ criticality of 
this BH in 
massive gravity in the extended {phase-space: in which the cosmological constant should 
be treated as a themodynamic variable} and its conjugate variable as a thermodynamic volume. 
{Moreover}, we derive the BH thermodynamic 
equation of state in analogy with the Van-der-Waal's equation of state. At the inflection 
point, we compute the critical pressure, the critical volume and the critical temperature. 
Using these constants, we compute the critical ratio. 
Moreover, we derive the reduced equation of state in terms of the reduced pressure, 
the reduced volume and the reduced temperature.

The organization of the paper is as follows. In Sec.~(\ref{cgg}), we have given the brief 
introduction of BHs in massive gravity in AdS space. In Sec.~(\ref{cgg1}), we have studied the 
critical behaviour  in extended phase space. In Sec.~(\ref{ehren}), we have described the 
Ehrenfest scheme for $P-V$ criticality in the extended phase space. 
Finally, we have given the conclusions in Sec.~(\ref{dis}).

\section{\label{cgg} A brief introduction of BHs in  Massive Gravity in AdS Space}
Here we would like to give a brief introduction of BHs in massive gravity in  the AdS space. In Einstein's 
general theory of relativity, the graviton has no mass but in {theory of} 
masive gravity the graviton has mass. That's 
why, it is called massive gravity. On the other hand, massive gravity 
is a kind of gravity which modified the Einstein's general theory of gravity.  
It is endowed with a mass which is called massive graviton. The idea of massive gravity  was first 
initiated  in 1939 by Fierz and Pauli~\cite{pauli}. 
In this theory, they proposed that a massive 2-spin propagates on a background of the flat spacetime.
They found that the graviton has five degrees of freedom. Afterwards, Boulware and Deser~\cite{deser} 
proved that Fierz-Pauli's prescription sustains from ghosts when one taking into account the non-linear 
extension. In 2010, de Rham, Gabadadze and Tolley proposed that a massive gravity theory which is a free 
from ghost i.e. ``ghost free gravity''. 
Now it is popularly known as dRGT theory in the literature. There exists other massive gravity models in the 
literature such as DGP model~\cite{dvali} and BH T model~\cite{townsend}. 
For a detail review on the ``Massive Gravity'' one could see the Ref. ~\cite{rham} and Ref.~\cite{hinter} 
[See also ~\cite{manos,hendi}].

What is interesting in the massive gravity is that it explains the ``acceleration of the universe'' which  
does not require any dark energy or any cosmological constant parameter. In 2016, in the first history of 
science the gravitional waves have been observed by the LIGO team in which the graviton mass is restricted 
to $m_{g}<1.2 \times 10^{-22} eV/c^2$ ~\cite{ligo16}.

There are several interesting theories of massive gravity. Among them one is a special theory of massive gravity
in which the Lorentz symmetry is broken which is popularly known as ``Lorentz symmetry breaking theory''. The 
reasons behind this symmetry breaking is that the spacetime contains a scalar field which is called Goldstone 
fields. Such fields are coupled to gravity through non-derivative coupling. When the Lorentz symmetry is broken 
spontaneously, the massive graviton gets the mass which is quite analogous to Higgs mechanism.

We do not repeat here  the derivation for this kind of BHs and the action for massive gravity theory
which could be found in the Ref.~\cite{bebrone11,commeli11,fernando17}. 
The metric~\cite{fernando17} for such type of BHs could be written as  
\begin{eqnarray}
ds^2 &= & -{\cal Y}(r) dt^2 + \frac{dr^2}{{\cal Y}(r)} +r^2 d\Omega_{2}^2~,~\label{mg}
\end{eqnarray}
where,
\begin{eqnarray}
{\cal Y}(r) &=& 1-\frac{2M}{r}-\gamma \frac{Q^2}{r^\lambda}-\frac{\Lambda}{3} r^2 .~\label{mg2}
\end{eqnarray}
and $d\Omega_{2}^2$ is the metric on unit sphere in two dimensions. Where $Q$ denotes a scalar charge related 
to massive gravity. The parameter $\gamma$ may have the values $\pm 1$ leads to different types of geometry 
and the value of the integration constant $\lambda>0$~\cite{fernando17}.

When $\gamma=1$, the BH solution is quite analogous to the Schwarzschild-AdS BH which has a single horizon. 
For the parameter $\lambda<1$, the function ${\cal Y}(r)$ ``dominates at large distances'' thus the ADM 
mass parameter become diverges. Whereas for $\lambda>1$, the BH solution becomes Schwarzschild-AdS type. 
Therefore throughout the work we will take the value of $\lambda>1$.

When $\gamma=-1$ and $\lambda=2$, one obtains the BH solution 
which is quite similar to the RN-AdS BH which has two horizons. 

\section{\label{cgg1} $P-V$ criticality of BHs in massive gravity in AdS space}
In this section, we would like to study the critical behaviour of BHs in massive 
gravity by using the extended phase-space formalism. Let us now put 
$-\frac{\Lambda}{3}=\frac{1}{\ell^2}$ for AdS case. Then the metric function becomes
\begin{eqnarray}
{\cal Y}(r) &=& 1-\frac{2M}{r}-\gamma \frac{Q^2}{r^\lambda}+\frac{r^2}{\ell^2} .~\label{mg3}
\end{eqnarray}
The outer horizon location can be determined from the condition ${\cal Y}(r_{+})=0$. 
The quantity $M$ denotes the ADM mass of the BH which could be identified as enthalpy 
of the system in extended phase-space. The parameter $Q$ denotes the total charge of 
the BH. 

Since we are dealing with extended phase-space in which the thermodynamic pressure~($P$) is equal to 
the negative cosmological constant~($\Lambda$) divided by $8\pi$~(where $G=c=k_{B}=\hslash$=1) i.e.
\begin{eqnarray} 
P &=& -\frac{\Lambda}{8\pi}=\frac{3}{8\pi \ell^2}  ~.\label{pmg}
\end{eqnarray}
and the corresponding thermodynamic volume can be written as
\begin{eqnarray}
V &=& \left(\frac{\partial M}{\partial P}\right)_{S,Q}  ~.\label{vmg}
\end{eqnarray} 
This volume for BHs in massive gravity in AdS space  is
\begin{eqnarray}
V &=& \frac{4}{3}\pi r_{+}^3  ~.\label{vmg2}
\end{eqnarray} 
where $r_{+}$ corresponds to the outer horizon radius.

The entropy of BH can be defined as 
\begin{eqnarray}
{\cal S}  &=& \frac{{\cal A}}{4} ~.\label{mg5}
\end{eqnarray}
where the area of BH is given by
\begin{eqnarray}
{\cal A} &=& 4\pi r_{+}^2 ~.\label{mg6}
\end{eqnarray}
Now the BH temperature is calculated to be 
\begin{eqnarray}
T &=& \frac{{\cal Y}'(r)}{4\pi}= 
\frac{1}{4\pi r_{+}} \left(1+8\pi P r_{+}^2+\frac{\gamma(\lambda-1)Q^2}{r_{+}^\lambda} \right)~.\label{mg7}
\end{eqnarray} 
The first law of thermodynamics in the extended phase-space becomes
\begin{eqnarray}
dM &=& T d{\cal S}+\Phi dQ+ V dP ~. \label{mg10}
\end{eqnarray}
where $\Phi$ is the electric potential given by 
\begin{eqnarray}
\Phi  &=&- \frac{\gamma Q} {r_{+}^{\lambda-1}} ~.\label{mg11}
\end{eqnarray}
{Thus the Smarr-Gibbs-Duhem relation~\cite{fer18} becomes }
\begin{eqnarray}
M &=& H= 2T {\cal S} - 2P V +\frac{\lambda}{2} Q \Phi ~. \label{mg12}
\end{eqnarray}
Now we are ready to derive the BH equation of state in the extended phase-space. Therefore we 
would like to determine the critical constants at the critical point. The critical point occurs  
at the point of inflection. Using Eq.~(\ref{mg7}), one could write the BH equation of state as
{
\begin{eqnarray}
P &=& \frac{T} {2r_{+}} -\frac{1}{8\pi r_{+}^2}-\frac{\gamma(\lambda-1)Q^2}{8\pi r_{+}^{\lambda+2}} ~. \label{mg15}
\end{eqnarray}
where $r_{+}=\left(\frac{3V}{4\pi}\right)^{1/3}$. In terms of specific volume $v=2r_{+}$ the above equation can be 
re-written as 
\begin{eqnarray}
P &=& \frac{T} {v} -\frac{1}{2\pi v^2}-\frac{2^{\lambda-1}\gamma(\lambda-1)Q^2}{\pi v^{\lambda+2}} ~.\label{mg16}
\end{eqnarray}
}
The critical point could be determined from the following conditions
\begin{eqnarray}
\frac{\partial P}{\partial v} &=& \frac{\partial^2 P}{\partial v^2}=0  ~. \label{mg18}
\end{eqnarray}
Using this criterion one obtains the critical values 
\begin{eqnarray}
P_{c} &=& \frac{\lambda}{2\pi(\lambda+2) [\gamma(1-\lambda) (1+\lambda)(2+\lambda)2^{\lambda-1}Q^2]^{\frac{2}{\lambda}}}~,\\
v_{c} &=&  [\gamma(1-\lambda) (1+\lambda)(2+\lambda)2^{\lambda-1}Q^2]^{\frac{1}{\lambda}}~, \\
T_{c} &=& \frac{\lambda}{\pi(\lambda+1) [\gamma(1-\lambda) (1+\lambda)(2+\lambda)2^{\lambda-1}Q^2]^{\frac{1}{\lambda}}} 
~. \label{mg19}
\end{eqnarray}
The critical thermodynamic volume could be obtain from the critical radius as
\begin{eqnarray}
V_{c} &=& \frac{4}{3}\pi r_{c}^3=\frac{\pi}{6}[\gamma(1-\lambda) (1+\lambda)(2+\lambda)2^{\lambda-1}Q^2]^{\frac{3}{\lambda}}
~. \label{mg20}
\end{eqnarray}
Now the critical ratio is computed to be 
\begin{eqnarray}
Z_{c} &=& \frac{P_{c} v_{c}}{T_{c}}=\frac{(1+\lambda)}{2(2+\lambda)} ~. \label{mg21}
\end{eqnarray}
This is a universal ratio. For RN-AdS BH, this {ratio is $\frac{3}{8}$}. 
Moreover, the Van-der-Waal's constants are 
\begin{eqnarray}
 a &=& \frac{9\lambda}{8\pi(1+\lambda)}~, \\
 b &=& \frac{[\gamma(1-\lambda) (1+\lambda)(2+\lambda)2^{\lambda-1}Q^2]^{\frac{1}{\lambda}}}{3} ~. \label{mg22}
\end{eqnarray}
It implies that the presence of the charge parameter reduces the effective volume smaller because of $b$ is proportional 
to $Q$.

Therefore ``the law of corresponding states'' could be written as   
\begin{eqnarray}
 2\left(2+\lambda \right)\Theta &=& \left(1+\lambda\right)\varPhi\left[\sigma+\left(1+\frac{2}{\lambda}\right)\frac{1}
 {\varPhi^2}\right]-\frac{2}{\lambda} \frac{1}{\varPhi^{\lambda+1}}
 ~. \label{mg24}
\end{eqnarray}
where $\Theta$, $\varPhi$ and $\sigma$ could be defined as 
\begin{eqnarray}
\Theta &=& \frac{T}{T_{c}},\,\, \varPhi = \frac{v}{v_{c}}, \,\, \sigma = \frac{P}{P_{c}} ~. \label{mg26}
\end{eqnarray}
and these quantities $\Theta$, $\varPhi$ and $\sigma$ are called the reduced temperature, the reduced volume and 
the reduced pressure respectively. Thus the above equation is called the \emph{reduced equation of state} or 
\emph{the law of corresponding states}.

Now we should compute the specific heat to determine the local thermodynamic stability of the BH. There are two 
specific heats in BH thermodynamics as well as in classical thermodynamics. Namely, the specific heat at constant 
volume and the specific heat at constant pressure. They are defined as 
\begin{eqnarray}
 C_{V}  &=& T \left(\frac{\partial S}{\partial T}\right)_{V} .~\label{cv}
\end{eqnarray}
and 
\begin{eqnarray}
 C_{P}  &=& T \left(\frac{\partial S}{\partial T}\right)_{P} .~\label{cp}
\end{eqnarray}
First, we should calculate the specific heat at constant 
volume. To determine it we first define the free energy which is calculated to be 
\begin{eqnarray}
 F &=& \frac{r_{+}}{2} \left[1-2\pi r_{+}T-\frac{\gamma Q^2}{r_{+}^\lambda}\right] ~. \label{mg27}
\end{eqnarray}
From the above equation we can easily compute the entropy which is defined to be 
\begin{eqnarray}
 S  &=& - \left(\frac{\partial F}{\partial T}\right)_{V} =\pi r_{+}^2.~\label{cvf}
\end{eqnarray}
Since the entropy $S$ is independent of $T$ thus one gets 
\begin{eqnarray}
C_{V} &=& 0  .~\label{cv1}
\end{eqnarray}
The temperature in terms of entropy and pressure could be defined as 
\begin{eqnarray}
T~(S,P) &=& \frac{1}{4\sqrt{\pi S}}\left[ 1+8PS+\frac{\gamma(\lambda-1)\pi^{\frac{\lambda}{2}}Q^2}
{S^{\frac{\lambda}{2}}}\right] .~\label{tsp}
\end{eqnarray}
To determine the local thermodynamic stability one should calculate the specific heat at constant 
pressure which is turned out to be
\begin{eqnarray}
C_{P} &=& 2S \frac{\left[S^\frac{\lambda}{2}+8PS^{1+\frac{\lambda}{2}}+\gamma(\lambda-1)\pi^\frac{\lambda}{2} Q^2\right]}
{\left[8PS^{1+\frac{\lambda}{2}}-S^\frac{\lambda}{2}-\gamma(\lambda-1)(\lambda+1)\pi^\frac{\lambda}{2} Q^2 \right]}  
.~\label{cp1}
\end{eqnarray}
For local thermodynamic stability it requires that $C_{P}>0$ and the specific heat $C_{P}$ diverges at 
\begin{eqnarray}
8PS^{1+\frac{\lambda}{2}}-S^\frac{\lambda}{2}-\gamma(\lambda-1)(\lambda+1)\pi^\frac{\lambda}{2} Q^2 &=& 0 
.~\label{cp2}
\end{eqnarray}
i.e. precisely at the critical point. It must be noted that the specific heat that $C_{P}$ but 
not $C_{V}$ diverges at the critical point. It could be observed from the above calculation. 

To determine the global thermodynamic stability of the BH one must compute the Gibbs free energy 
which is determined to be
\begin{eqnarray}
G &=& G(T,P)=\frac{r_{+}}{4}\left[1-\frac{8\pi P}{3}r_{+}^2-\frac{\gamma(\lambda+1)Q^2}{r_{+}^\lambda}\right].
~\label{gf}
\end{eqnarray}
where $r_{+}$ is a function of temperature and pressure, $r_{+}=r_{+}(P,T)$ by using the equation of state
~Eq.~(\ref{mg15}).  { In terms of BH temperature the above equation can be re-expressed 
as 
\begin{eqnarray}
G &=& \frac{r_{+}}{3}\left[1-\pi r_{+} T-\frac{\gamma(\lambda+2)Q^2}{2 r_{+}^\lambda}\right].
~\label{gff}
\end{eqnarray}
Solving the equation G=0, one obtains the critical temperature 
\begin{eqnarray}
T_{c} &=& \frac{1}{\pi r_{+}^{\lambda+1}} \left[r_{+}^{\lambda}
-\frac{\gamma(\lambda+2)Q^2}{2}\right] ~\label{gf1}
\end{eqnarray}
The value of $r_{+}$ can be eliminated by using  Eq.~(\ref{mg7}). Although it is a very difficult 
task for this BH but it is easy for $Q=0$. In this case $r_{+}=\sqrt{\frac{3}{8\pi P}}$ and 
$T_{c}=T_{HP}=\sqrt{\frac{8P}{3\pi}}$. This is called the Hawking-Page phase transition temperature. 
For $T>T_{c}$, the BH will be stable one where the Gibb's free energy is minimum. The Gibb's free 
energy has been studied in detail in Ref.~\cite{fernando16}.}

\section{\label{ehren} Ehrenfest equations for BHs in massive gravity in the Extended Phase Space}
Phase transition is a fascinatic topic of research in BH thermodynamics. In order to 
understand the nature of phase transitions whether it is first order phase transition or second 
order phase transition at the critical point in the extended phase space we should examine the 
Clausius-Clapeyron equations 
and Ehrenfest equations~\cite{prig}. If the Clausius-Clapeyron equations are satisfied that means 
there should exist the first order phase transition and if the Ehrenfest equations are satisfied 
that implies there should exist second order phase transition. In this section we would try to 
understand what type of phase transition should occur for BHs in massive gravity? 

For analytical check of Ehrenfest equations for BHs in massive gravity in the extended phase space 
first we should consider the comparison between BH thermodynamic parameters with the classical 
thermodynamic parameters i.e.
\begin{eqnarray}
 V &=& Q  \\
 P &=& -\Phi
\end{eqnarray}
This nice comparison of Ehrenfest equations  first introduced in BH thermodynamics by 
Banerjee et al.~\cite{banerjee,mo} as 
\begin{eqnarray}
-\left(\frac{\partial \Phi}{\partial T} \right)_{S} &=& \frac{{C_{\Phi}}_{2}-{C_{\Phi}}_{1}}
{TQ(\alpha_{2}-\alpha_{1})}=\frac{\Delta C_{\Phi}}{T Q\Delta \alpha}  .~\label{eh1} \\
-\left(\frac{\partial \Phi}{\partial T} \right)_{Q} &=& \frac{\alpha_{2}-\alpha_{1}}
{{K_{T}}_{2}-{K_{T}}_{1}}=\frac{\Delta \alpha}{\Delta K_{T} }  .~\label{eh2}
\end{eqnarray}
where $\alpha$ and $K_{T}$ are defined as 
\begin{eqnarray}
 \alpha &=& \frac{1}{Q} \left(\frac{\partial Q}{\partial T} \right)_{\Phi}  \\
 K_{T} &=&  \frac{1}{Q} \left(\frac{\partial Q}{\partial \Phi} \right)_{T}
\end{eqnarray}
which are analogous to the volume expansion coefficient and the isothermal compressibility.

The Ehrenfest equations in classical thermodynamics which could be found in any standard text 
book of thermodynamics are 
\begin{eqnarray}
\left(\frac{\partial P}{\partial T} \right)_{S} &=& \frac{{C_{P}}_{2}-{C_{P}}_{1}}
{TV(\alpha_{2}-\alpha_{1})}=\frac{\Delta C_{P}}{T V\Delta \alpha}  .~\label{eh3} \\
\left(\frac{\partial P}{\partial T} \right)_{V} &=& \frac{\alpha_{2}-\alpha_{1}}
{{K_{T}}_{2}-{K_{T}}_{1}}=\frac{\Delta \alpha}{\Delta K_{T} }  .~\label{eh4}
\end{eqnarray}
where $\alpha$ and $K_{T}$ are
\begin{eqnarray}
\alpha &=& \frac{1}{V} \left(\frac{\partial V}{\partial T} \right)_{P}  \\
K_{T}  &=& - \frac{1}{V} \left(\frac{\partial V}{\partial P} \right)_{T}
\end{eqnarray} 
$\alpha$ is the volume expansion coefficient and $K_{T}$ is the isothermal 
compressibility coefficient.  Now our motivation here is that the above classical 
Ehrenfest equations could be implimented in the BH thermodynamics particularly in the extended 
phase space because this formalism gives us the specific heat of BHs at constant 
pressure. 

Now we are ready to compute the relevant thermodynamic quantities in the extended phase 
space for massive gravity. First we need to compute the BH temperature which is found to 
be from Eq.~(\ref{tsp})
\begin{eqnarray}
T &=& \frac{1}{4\sqrt{\pi S}}\left[ 1+8PS+\frac{\gamma(\lambda-1)\pi^{\frac{\lambda}{2}}Q^2}
{S^{\frac{\lambda}{2}}}\right] .~\label{tsp1}
\end{eqnarray}
The other relevant quantities are 
\begin{eqnarray}
C_{P} &=& T \left(\frac{\partial S}{\partial T} \right)_{P} 
= 2S \frac{\left[S^\frac{\lambda}{2}+8PS^{1+\frac{\lambda}{2}}+\gamma(\lambda-1)\pi^\frac{\lambda}{2} Q^2\right]}
{\left[8PS^{1+\frac{\lambda}{2}}-S^\frac{\lambda}{2}-\gamma(\lambda-1)(\lambda+1)\pi^\frac{\lambda}{2} Q^2 \right]}  
.~\label{cp3} \\
\alpha &=& \frac{1}{V} \left(\frac{\partial V}{\partial T} \right)_{P} =\frac{12 \sqrt{\pi}S^(\frac{\lambda+1}{2})}
{\left[8PS^{1+\frac{\lambda}{2}}-S^\frac{\lambda}{2}-\gamma(\lambda-1)(\lambda+1)\pi^\frac{\lambda}{2} Q^2 \right]}
.~\label{cp4}
\end{eqnarray}
Now using the following identity
\begin{eqnarray}
\left(\frac{\partial P}{\partial V} \right)_{T} \left(\frac{\partial V}{\partial T} \right)_{P} 
\left(\frac{\partial T}{\partial P} \right)_{V} &=& -1 .~\label{cp5}
\end{eqnarray}
one obtains the expression for the coefficient of isothermal compressibility as 
\begin{eqnarray}
K_{T}  &=& - \frac{1}{V} \left(\frac{\partial V}{\partial P} \right)_{T}=
\frac{24 S^{1+\frac{\lambda}{2}}}
{\left[8PS^{1+\frac{\lambda}{2}}-S^\frac{\lambda}{2}-\gamma(\lambda-1)(\lambda+1)\pi^\frac{\lambda}{2} Q^2 \right]}
.~\label{cp6}
\end{eqnarray} 
From the above calculation we observed that the denominator in the expressions of $C_{P}$, $\alpha$ and $K_{T}$ are 
same. It indicates that the volume expansion coefficient and isothermal compressibility coefficient are singular at 
the critical point as the $C_{P}$ does.  

Now we would like to check the validity of the Ehrenfest equations at the critical point. Using the expression of 
volume expansion one could find 
\begin{eqnarray}
\alpha V &=& \left(\frac{\partial V}{\partial T} \right)_{P} 
=\frac{C_{P}}{T} \left(\frac{\partial V}{\partial S} \right)_{P}  .~\label{cp7}
\end{eqnarray} 
Thus the  R.H.S of Eq.~(\ref{eh3}) at the critical point becomes
\begin{eqnarray}
\frac{\Delta C_{P}}{TV\Delta\alpha} &=& \left[\left(\frac{\partial S}{\partial V} \right)_{P} \right]_{c} 
.~\label{cp8}
\end{eqnarray} 
which is computed to be 
\begin{eqnarray}
\frac{\Delta C_{P}}{TV\Delta\alpha} &=& \frac{\sqrt{\pi}}{2\sqrt{S_{c}}} .~\label{cp9}
\end{eqnarray} 
Interestingly this is independent of both the parameters $\gamma$ and $\lambda$. Using Eq.~(\ref{tsp1}), one 
could compute the L.H.S of Eq.~(\ref{eh3}) as 
\begin{eqnarray}
\left[\left(\frac{\partial P}{\partial T} \right)_{V} \right]_{c} &=&  \frac{\sqrt{\pi}}{2\sqrt{S_{c}}}
.~\label{cp10}
\end{eqnarray} 
From the Eq.~(\ref{cp9}) and the Eq.~(\ref{cp10}) we observed that the first Ehrenfest equation is 
satisfied at the critical point in the extended phase space for massive gravity.

Analogously using the volume expansion coefficient and isothermal compressibility coefficient, one 
obtains 
\begin{eqnarray}
\frac{\Delta \alpha}{\Delta K_{T}} &=& \left[\left(\frac{\partial P}{\partial T} \right)_{V} \right]_{c} =
\frac{\sqrt{\pi}}{2\sqrt{S_{c}}}
.~\label{cp11}
\end{eqnarray} 
This implies that the second Ehrenfest equation is satisfied at the critical point. 

From the above calculation we can conclude that both Ehrenfest equations are satisfied at the 
critical point in the extended phase space for massive gravity. Now we compute the Prigogine-Defay~(PD) 
ratio using Eq.~(\ref{cp8}) and  Eq.~(\ref{cp11}) as 
\begin{eqnarray}
\varPi &=& \frac{\Delta C_{P}\Delta K_{T} }{TV(\Delta\alpha)^2} = 1 .~\label{cp12}
\end{eqnarray} 
The Prigogine-Defay~\cite{prig} ratio~\footnote{This ratio was first suggested by 
Prigogine and Defay. That's why it is called PD ratio.}
$\varPi=1$ and Ehrenfest equations are satisfied means that the 
phase transition in the extended phase space for massive gravity is a second order one. This is an 
interesting result for this BH. This result is also compatible with the nature of liquid-gas 
phase transition in the extended phase space and at the critical point. Moreover, this result is 
strengthening the analogy between the massive gravity and the liquid-gas system.

\section{\label{dis} Conclusion}
We demonstrated the extended phase space thermodynamics of BHs in 
massive gravity. Particularly, we have examined the critical behaviour for this BH 
by using the extended phase space procedure. In this procedure one may treated the cosmological 
constant as a dynamical pressure and its conjugate variable as a thermodynamic volume.
We derived the BH entropy, the BH temperature, the BH equation 
of state, the first law of thermodynamics, the Smarr formula, the Gibbs free energy and 
finally  the law of corresponding states. We have found that the complete BH 
thermodynamic system which is quite analogous to the liquid-gas system.

We also derived the specific heat at constant pressure, the coefficient of volume expansion and the coefficient 
of isothermal compressibility. Moreover we derived the Pregogine-Defay ratio and we showed that the value of this 
ratio is equal to one. Which {indicates}  that the phase transition is second order one. 
{Furthermore using the 
analogy between the thermodynamic state variables and the BH parameters i.e. $V\leftrightarrow Q$ and 
$P\leftrightarrow \Phi$, we showed that in the extended phase space the Ehrenfest equations are satisfied for 
this BH}. The consequences of this study gives us further  very intense understanding between 
the BH of massive gravity in the extended phase space and the liquid-gas systems.

\end{document}